\PassOptionsToPackage{colorlinks=true, allcolors=blue}{hyperref}
\documentclass[pdflatex,sn-mathphys-num]{sn-jnl}
\usepackage[utf8]{inputenc}
\usepackage{multirow}%
\usepackage{mathrsfs}%
\usepackage{textcomp}%
\usepackage{booktabs}%
\usepackage{listings}%
\usepackage{natbib}
\usepackage{amsmath, amssymb, amsfonts}
\usepackage{graphicx}
\usepackage{booktabs}
\graphicspath{{Figures/}} 
\usepackage{xcolor}
\usepackage{bm}
\usepackage[colorlinks=true, allcolors=blue]{hyperref}
\usepackage{siunitx}
\usepackage{comment}
\usepackage{afterpage}
\usepackage{caption}
\usepackage{setspace}

\begin{document}

\title{PySlyde: A Lightweight, Open-Source Toolkit for Pathology Preprocessing}

\author[1,2,*]{\fnm{Gregory}\sur{Verghese}}\email{gregory.e.verghese@kcl.ac.uk}
\author[2,3,*]{\fnm{Anthony} \sur{Baptista}}\email{anthony.baptista@kcl.ac.uk}
\author[2,*]{\fnm{Chima} \sur{Eke}}\email{chima.eke@kcl.ac.uk}
\author[2,*]{\fnm{Holly} \sur{Rafique}}\email{holly.rafique@kcl.ac.uk}
\author[1]{\fnm{Mengyuan} \sur{Li}}
\author[2]{\fnm{Fathima} \sur{Mohamed}}
\author[2,4]{\fnm{Ananya} \sur{Bhalla}}
\author[2]{\fnm{Lucy} \sur{Ryan}}
\author[1,2]{\fnm{Michael} \sur{Pitcher}}
\author[1,2]{\fnm{Enrico} \sur{Parisini}}
\author[1,5]{\fnm{Concetta} \sur{Piazzese}}
\author[1,6]{\fnm{Liz} \sur{Ing-Simmons}}
\author[1,2]{\fnm{Anita} \sur{Grigoriadis}}\email{anita.grigoriadis@kcl.ac.uk}

\affil[1]{\orgdiv{PharosAI}, \orgname{King’s College London}, \orgaddress{\city{London}, \postcode{WC2R 2LS}, \country{UK}}}
\affil[2]{\orgdiv{School of Cancer and Pharmaceutical Sciences, Faculty of Life Sciences and Medicine}, \orgname{King’s College London}, \orgaddress{\city{London}, \postcode{WC2R 2LS}, \country{UK}}}
\affil[3]{\orgdiv{The Alan Turing Institute}, \orgname{The British Library}, \orgaddress{\city{London}, \postcode{NW1 2DB}, \country{UK}}}
\affil[4]{\orgdiv{The Francis Crick Institute}, \orgaddress{\city{London}, \postcode{NW1 1AT}, \country{UK}}}
\affil[5]{\orgdiv{Barts Life Sciences}, \orgname{Barts Health NHS Trust}, \orgaddress{\city{London}, \postcode{E1 1BB}, \country{UK}}}
\affil[6]{\orgdiv{eResearch}, \orgname{King's College London}, \orgaddress{\city{London}, \postcode{WC2R 2LS}, \country{UK}}}

\affil[*]{These authors contributed equally}

\abstract{
The integration of artificial intelligence (AI) into pathology is advancing precision medicine by improving diagnosis, treatment planning, and patient outcomes. Digitised whole-slide images (WSIs) capture rich spatial and morphological information vital for understanding disease biology, yet their gigapixel scale and variability pose major challenges for standardisation and analysis. Robust preprocessing, covering tissue detection, tessellation, stain normalisation, and annotation parsing is critical but often limited by fragmented and inconsistent workflows. We present PySlyde, a lightweight, open-source Python toolkit built on OpenSlide to simplify and standardise WSI preprocessing. PySlyde provides an intuitive API for slide loading, annotation management, tissue detection, tiling, and feature extraction, compatible with modern pathology foundation models. By unifying these processes, it streamlines WSI preprocessing, enhances reproducibility, and accelerates the generation of AI-ready datasets, enabling researchers to focus on model development and downstream analysis.
}

\maketitle

\section*{Summary}
\hfill

The integration of artificial intelligence (AI) into pathology represents a transformative opportunity for precision medicine, with the potential to radically improve diagnosis, treatment planning, and patient outcomes~\cite{Verghese2023ComputationalPI,Song2023ArtificialIF}. Advances in computational pathology are enabling the analysis of complex histopathological data at unprecedented scale and resolution. By combining modern computer vision, machine learning, and pathology, researchers can extract quantitative biomarkers, detect subtle morphological patterns, and develop predictive models that complement human expertise~\cite{vanderLaak2021DeepLI}. \\

Central to these developments is the digitisation of histopathology slides into whole slide images (WSIs). WSIs offer valuable spatial and morphological insights that can deepen our understanding of cancer biology. However, their size (gigapixel-scale), complexity, and variability introduce significant computational and standardisation challenges for downstream computational tasks~\cite{McGenity2023ArtificialII}. Preprocessing WSIs, such as tissue detection, tessellation (splitting into smaller tiled images), stain normalisation and annotation parsing are essential steps before any downstream AI analysis can occur~\cite{Pocock2022TIAToolboxAA}. Building robust, reproducible pipelines for WSI preprocessing is fundamental to computational pathology research. However, existing workflows often rely on ad hoc proprietary scripts, fragmented tools, and inconsistent formats, making standardisation and reproducibility difficult. \\

PySlyde is a lightweight, open-source Python toolkit wrapped around OpenSlide~\cite{GOODE201327}, designed to address this gap and provide an intuitive and efficient approach to quickly preprocess WSIs. It provides a simple API to perform steps such as WSI loading, annotation handling, tissue detection, tile generation, and feature extraction with support for the latest pathology foundation models. PySlyde aims to lower the barrier to entry for pathology researchers, standardise preprocessing workflows, and accelerate the development of AI-ready datasets for computational pathology research, enabling researchers to focus on model development and other downstream tasks.

\begin{figure}[!hb]
    \centering
    \includegraphics[width=1\textwidth]{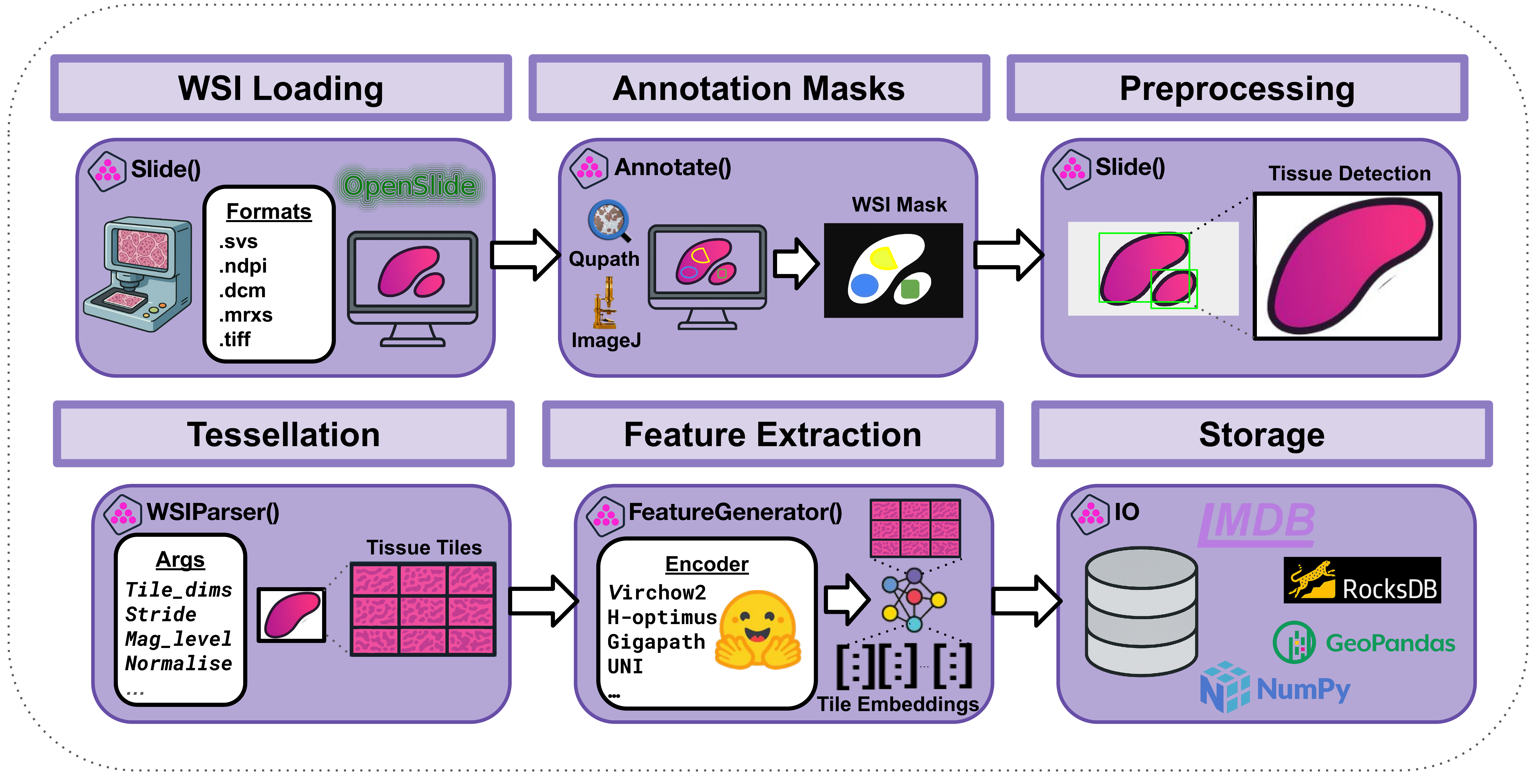}
    \caption{Schematic overview of the WSI processing workflow supported by PySlyde. The overview is demonstrating a sample of functionality across the main classes and modules \texttt{Slide}, \texttt{WSIParser}, \texttt{FeatureGenerator} and the \texttt{IO}. Top row, from left to right: WSI loading using compatible formats through OpenSlide; annotation mask generation with support for QuPath or ImageJ formats; and preprocessing steps such as tissue detection. Bottom row, from left to right: Tessellation of WSIs into tissue tiles using the \texttt{WSIParser} class; feature extraction from tiles with support for pretrained encoders such as Virchow2, H-optimus, Gigapath, or UNI; and structured storage of tile embeddings and metadata using RocksDB, NumPy, or LMDB.}
    \label{fig:example}
\end{figure}

\section*{Statement of Need}
\hfill

WSI preprocessing is a foundational but time-consuming step in computational pathology. Common tasks, such as annotation parsing, binary and annotation mask generation, stain normalisation, and tessellation often involve custom scripts, multiple packages and ad hoc pipelines built for a single dataset or project. This fragmentation hinders reproducibility, increases technical debt, and slows the development and translation of AI methods into clinical workflows. The lack of standardised preprocessing practices thus remains one of the key barriers preventing the seamless integration of AI into diagnostic pathology~\cite{bilal2025foundationmodelscomputationalpathology}. \\

While powerful frameworks exist for model training and inference, few provide a simple, consistent interface for preparing WSIs at scale. PySlyde directly addresses this unmet need by providing a modular, well-documented, and extensible preprocessing toolkit that integrates seamlessly with existing ecosystems, including OpenSlide, NumPy, PyTorch, and modern foundation models enabling researchers to build reproducible, scalable pipelines with minimal code.

\section*{Features}
\hfill

PySlyde provides several core modules that streamline preprocessing of WSIs. The toolkit is structured around four main components: the \texttt{Slide} class, the \texttt{WSIParser} class, the \texttt{FeatureGenerator} class, and the \texttt{IO} module.

\begin{itemize}
  \item \textbf{The \texttt{Slide} Class:} Provides core functionality for working with individual WSIs.
  \begin{itemize}
    \item \textbf{WSI Loading:} Built on OpenSlide, PySlyde supports reading and processing of any OpenSlide-compatible WSI format.
    \item \textbf{Tissue Detection:} Implements tissue segmentation to create binary masks for filtering and selecting relevant regions.
    \item \textbf{Artifact Detection:} Integrates with HistoQC for detecting common WSI artifacts such as blurring, pen marks, and background noise.
    \item \textbf{Annotation Parsing:} Supports annotation files from QuPath~\cite{Bankhead2017QuPathOS} and ImageJ, and formats including JSON and CSV.
    \item \textbf{Mask Generation:} Enables fast segmentation mask creation and label extraction from annotations.
  \end{itemize}

  \item \textbf{The \texttt{WSIParser} Class:} Handles slide-level tiling and region management.
  \begin{itemize}
    \item \textbf{Region Extraction:} Supports bounding-box extraction and spatial filtering of tissue regions.
    \item \textbf{Efficient Tiling:} Performs configurable, memory-efficient tiling at user-defined magnifications and strides. Tiles and features can be exported as image files or to low-latency key-value stores such as LMDB or RocksDB.
    \item \textbf{Stain Normalisation:} Includes methods such as Macenko~\cite{Macenko2009AMF}, Reinhard~\cite{reinhard2001}, and Vahadane~\cite{Vahadane2016StructurePreservingCN} for colour normalisation across tiles, enabling consistent downstream analysis.
  \end{itemize}

  \item \textbf{The \texttt{FeatureGenerator} Class:} Provides functionality for feature extraction using foundation models.
  \begin{itemize}
    \item \textbf{Feature Embedding:} Supports multiple pathology foundation models, including H-optimus-0~\cite{hoptimus2024}, cTransPath\cite{Wang2022TransformerbasedUC}, Virchow2\cite{Zimmermann2024Virchow2SS} and UNI2~\cite{Chen2024TowardsAG}, for computing  tile-level or slide-level embeddings for machine learning and clustering.
  \end{itemize}

  \item \textbf{The \texttt{IO} Module:} Manages data export, integration, and reproducibility.
  \begin{itemize}
    \item \textbf{Reproducibility:} Comprehensive documentation, quick-start examples, and API references support transparent and reusable workflows.
  \end{itemize}
\end{itemize}

\section*{Maintenance and Availability}
\hfill

PySlyde is actively maintained by contributors from PharosAI and King's College London. The project is hosted on GitHub under an Apache 2.0 license, with Python unit tests and community issue tracking. Source code, installation instructions, and documentation are freely available at \href{https://gregoryverghese.github.io/PySlyde/}{https://gregoryverghese.github.io/PySlyde/}. Users and contributors are encouraged to submit pull requests, report issues, and extend the toolkit to support emerging pathology and biomedical imaging standards.

\section*{Acknowledgements}
\hfill

We thank members of the King's College London Cancer Bioinformatics team, the PharosAI group, and collaborators at The Alan Turing Institute and Barts Life Sciences for feedback, testing, and support.  Holly Rafique, Pre-Doctoral Fellow NIHR303406, is funded by the NIHR, Chima Eke by CRUK and Anita Grigoriadis, and Gregory Verghese by Department of Science Innovation and Technology under PharosAI for this research project. Anthony Baptista, and Anita Grigoriadis acknowledge support from the CRUK City of London Centre Award [CTRQQR-2021/100004]. The views expressed in this publication are those of the author(s) and not necessarily those of the NIHR.


\end{document}